\documentclass[usenatbib,usegraphics]{mn2e}                 %

\usepackage{graphicx}
\usepackage{aas_macros}
\usepackage{natbib}

\def\kms{km\,s$^{-1}$}
\def\kmsMpc{km\,s$^{-1}$ Mpc$^{-1}$}
\def\Ha{H$\alpha$}

\def\psec{\ensuremath{\,.\!\!^{s}}}

\def\parcsec{\ensuremath{\,.\!\!\arcsec}}
\def\HeI{He\,{\sc i}}
\def\FeII{Fe\,{\sc ii}}
\def\Nifs{$^{56}$Ni}
\def\Cofs{$^{56}$Co}

\title[Supernova 2001gh]{Explosion of a massive, He-rich star at z=0.16}
\author[Elias-Rosa et al.]
{N. Elias-Rosa$^{1}$\thanks{E-mail: nancy.elias@oapd.inaf.it},
A. Pastorello$^{1}$, M. Nicholl$^{2}$, S. Valenti$^{3,4}$, S. Benetti$^{1}$, E. Cappellaro$^{1}$, \and
M. Turatto$^{1}$, G. Altavilla$^{5}$, M. T. Botticella$^{6}$, L.~A.~G.Monard$^{7}$, M. Riello$^{8}$, L. Zampieri$^{1}$ 
\\
$^1$INAF - Osservatorio Astronomico di Padova, vicolo dell'Osservatorio 5, I-35122 Padova, Italy. \\
$^2$Astrophysics Research Centre, School of Mathematics and Physics, Queen's University Belfast, Belfast BT7 1NN, UK\\
$^3$Las Cumbres Observatory Global Telescope Network, 6740 Cor tona Dr., Suite 102, Goleta, CA 93117, USA\\
$^4$Department of Physics, University of California, Santa Barbara, Broida Hall, Mail Code 9530, Santa Barbara, CA 93196-9530, USA\\
$^5$INAF - Osservatorio Astronomico di Bologna, V. Ranzani 1, I-40127 Bologna, Italy. \\
$^6$INAF - Osservatorio Astronomico di Capodimonte, Salita Moiariello, 16 80131 Napoli, Italy\\
$^7$Bronberg Observatory, Centre for Backyard Astrophysics (Pretoria), 0056 Tiegerpoort, South Africa\\
$^8$Institute of Astronomy, University of Cambridge, Madingley Road, Cambridge CB3 0HA, UK \\
}

\begin{document}

\date{Accepted ???. Received ???}

\pagerange{\pageref{firstpage}--\pageref{lastpage}} \pubyear{2014}

\maketitle

\label{firstpage}

\begin{abstract}

We present spectroscopic and photometric data of the peculiar SN~2001gh, discovered by the {\it Southern inTermediate Redshift ESO Supernova Search} (STRESS) at a redshift z=0.16. SN~2001gh has relatively high luminosity at maximum (M$_B$ = -18.55 mag), while the light curve shows a broad peak. An early-time spectrum shows an almost featureless, blue continuum with a few weak and shallow P-Cygni lines that we attribute to \HeI. \HeI\ lines remain the only spectral features visible in a subsequent spectrum, obtained one month later. A remarkable property of SN~2001gh is the lack of significant spectral evolution over the temporal window of nearly one month separating the two spectra. 
In order to explain the properties of SN~2001gh, three powering mechanism are explored, including radioactive decays of a moderately large amount of $^{56}$Ni, magnetar spin-down, and interaction of SN ejecta with circumstellar medium. We favour the latter scenario,
with a SN Ib wrapped in a dense, circumstellar shell. The fact that no models provide an excellent fit with observations, confirms the troublesome interpretation of the nature of SN~2001gh. 
A rate estimate for SN~2001gh-like event is also provided, confirming the intrinsic rarity of these objects.

\end{abstract}

\begin{keywords}
supernovae: general -- supernovae: individual:
SN 2001gh, SN 2002hy
\end{keywords}

%
%
\section{Introduction}\label{intro}
The most massive stars (larger than 25-30 M$_\odot$) are expected to lose their external hydrogen envelope in the final stages of their lives, becoming He-rich Wolf-Rayet stars (hereafter WRs) or, if they lose also their He mantle, C-O WRs. For decades, these stars were thought to explode as Type Ib or Ic supernovae (SNe), although there has been no direct detection of their progenitors to date. 

H depleted core-collapse (CC-) SNe display a variety of observed properties (\citealt{turatto07}) depending on the stellar parameters (e.g. mass, radius, angular momentum) and the configuration of their circumstellar environment (chemical composition, density, geometrical distribution). Occasionally, SNe Ib/c display luminous light curve peaks and broad spectral features, which are indicative of high expansion velocity of the ejecta. They are labelled as hypernovae, and became very popular since late 90's for their connection with gamma-ray bursts (GRBs; see \citealt{galama98,patat01,dellavalle07,nomoto07} for reviews). There are also intermediate cases such as the Type Ib SN~2005bf, characterized by a peculiar, double-peaked light curve, that appear as transitional events between canonical SN Ib/c and gamma-ray bursters (\citealt{anupama05,tominaga05,folatelli06}).

Even more rare, a few extremely luminous SNe Ib/c were proposed to synthesize a large quantity of \Nifs, e.g. the peculiar Type Ic SN~1999as, for which 4 M$_\odot$ \citep{deng01} of \Nifs\ were estimated. Because of their large inferred \Nifs\ masses, objects like SN 1999as and SN 2007bi \citep{galyam09,young10} were proposed to be pair-instability SNe \citep{galyam12}, although other plausible scenarios such as that driven by the radioactive decay, were also suggested (e.g. \citealt{moriya10,nicholl13}). 

The operations of modern transient searches bring to the discovery of new types of objects that are challenging the classical SN explosion scenarios. Super-luminous stripped envelope/H-poor SNe (e.g. \citealt{quimby11}, \citealt{galyam12}, and \citealt{inserra13}) are an example of newly discovered transients, having luminosities exceeding $10^{44}$ erg s$^{-1}$, slow-rising and bell-shaped light curves without a late-time tail declining following the $^{56}$Co decay rate, and blue spectra. A few alternative scenarios have been proposed to explain these events, for example those where the SNe luminosity has been powered by the spin-down of a rapidly rotating young magnetar \citep{kasen10}, or by conversion of kinetic energy into radiation following the shock of shells with different velocities. The latter phenomena may result from many different evolutionary scenarios including non-disruptive stellar explosions driven by pulsational pair instability (see e.g. \citealt{woosley07}), or by interaction of SN ejecta with a massive H- and He-poor circumstellar medium (CSM, e.g. \citealt{blinnikov10,chevalier11,moriya13}) or with more H-rich CSM, as in the case of the exceptionally luminous CSS121015:004244+132827 \citep{benetti14}. The same mechanism is invoked for the family of Type Ibn SNe (see e.g. \citealt{pastorello08}), whose spectra are characterized by broad metal lines with superposed narrower \HeI\ features, and are believed to be core-collapse SNe exploded within a He-rich CSM (e.g. \citealt{matheson00,pastorello08}). The prototype of this SN family is SN~2006jc \citep{pastorello07,foley07}, whose progenitor, likely a Wolf-Rayet star, was observed in outburst shortly before the stellar core-collapse.

Despite the rolling surveys have discovered thousands of transients and many new SN types, a mysterious object discovered more than a decade ago still remains an unique case. The peculiar SN~2001gh was discovered by the {\it Southern inTermediate Redshift ESO Supernova Search} (STRESS, \citealt{botticella08}; {\it ESO} stands for European Southern Observatory). The search was specifically designed to measure the rate of Type Ia and CC SNe at intermediate redshift. 

In the next section we will introduce the transient discovery. In section \ref{obs}, we will present the spectroscopic (section \ref{spec}) and photometric data (section \ref{ph}) of SN~2001gh, while in section \ref{comp1bpec} we will briefly compare the case of SN~2001gh with other peculiar Type Ib SNe. In section \ref{constr} we will attempt to constrain via data modeling the explosion and ejecta parameters of SN 2001gh and, hence, characterize the progenitor star. Finally, in section \ref{sum}, we will summarize our results. Additionally, an appendix has been included to illustrate previously unpublished data of a peculiar comparison object, the Type Ib SN 2002hy (Appendix \ref{sn02hy}).

%
%
\section{SN~2001gh}
\label{sn01ghdata}

\begin{figure}
\centering
\includegraphics[width=0.9\columnwidth]{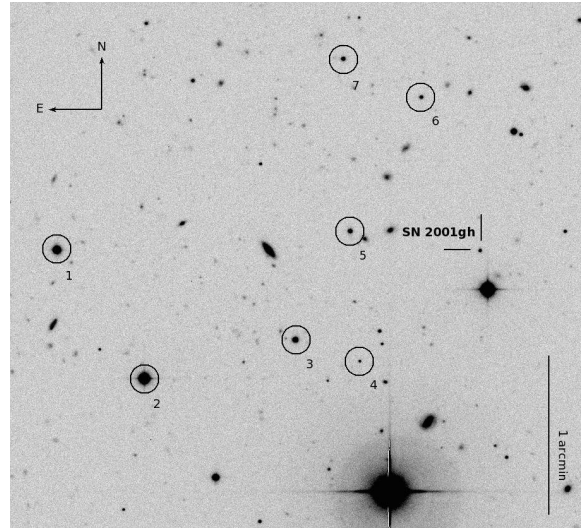} 
\caption{$V$-band image of SN~2001gh field taken with ESO 2.2 m + WFI on 2001 November 12. The stars used for the photometry calibration of the SN are labelled with numbers.}
\label{fig_FC01gh}
\end{figure}

SN~2001gh was discovered by \cite{altavilla01} on 2001 November 12.19 UT (UT dates will be used throughout this paper). The object was in the {\it STRESS} field labelled J1888 \citep{botticella08} at $\alpha = 00^{h}57^{m}03\psec63$, $\delta = -27^{o}42'32\parcsec90$ (J2000.0; see Figure \ref{fig_FC01gh}). SN~2001gh had a blue, featureless spectrum and was tentatively classified a Type II supernova at very early phases \citep{altavilla01}, although this classification has later been revised \cite{eliasrosa09}.

As we will see in the next section, the redshift estimated for this SN is z=0.16 $\pm$ 0.01. We will adopt a distance modulus of 39.33 $\pm$ 0.10 mag, as derived from the Robertson-Walker metric considering {\it z}, H$_{0}$ = 73 \kmsMpc, $\Omega_H$ = 0.27 and $\Omega_{\Lambda}$ = 0.73. We also considered a Galactic reddening correction of $E(B-V) = 0.016$ mag (\citealt{schlafly11}). As there is no spectroscopic signature of interstellar Na ID at the host galaxy redshift, the contribution of the host galaxy to the total reddening is assumed to be negligible. 

Unfortunately we cannot well constrain the explosion epoch of SN~2001gh. Given the strategy of STRESS, each field was observed on average every $\sim$ 3 months. In particular, our object was not detected on 2001 September 26, while the first detection is dated 2001 November 12, giving us a window of approximately 47 days in the explosion time. Considering the epoch adopted for the $B$ maximum light (see section \ref{ph}) and an average value of 17.5 days for the rise of typical Type Ib SNe \citep{valenti08a}, the explosion date for SN~2001gh could have happened just two days before its discovery, on JD = 2452225.

%
%
\section{Observational analysis}
\label{obs}

\subsection{Spectroscopy}
\label{spec}

\begin{figure*}
\centering
\includegraphics[width=1.8\columnwidth]{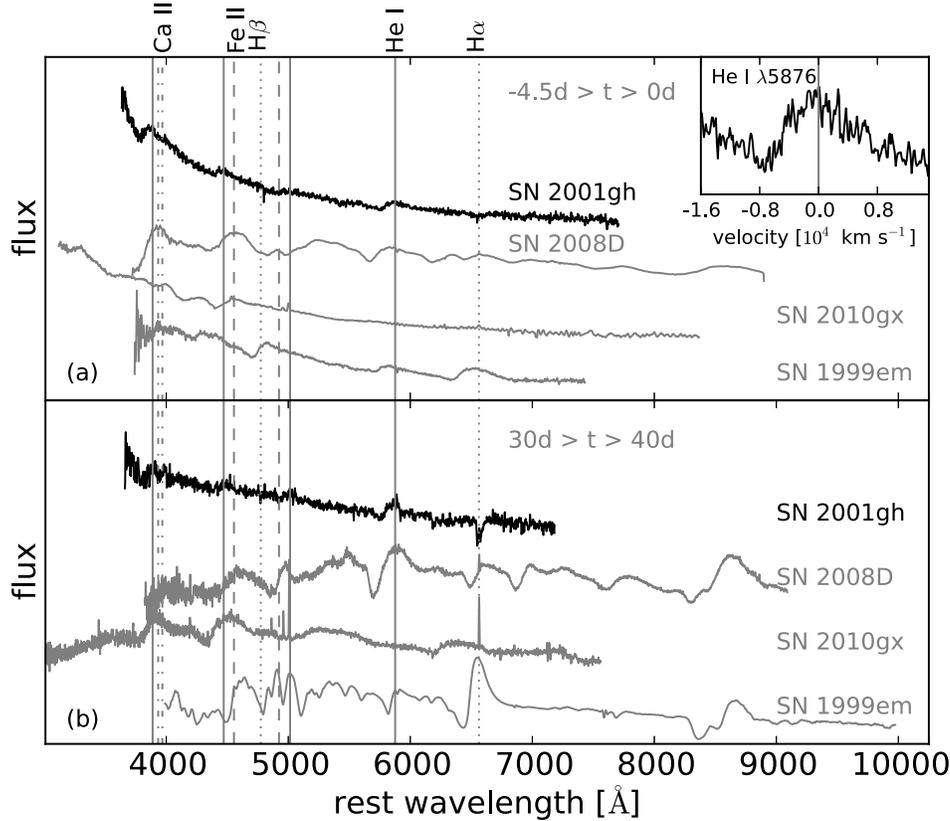}
\caption{Comparison between optical spectra of SN~2001gh a few days before maximum light (\textit{a}) and one month later (\textit{b}) with those of SN~Ib~2008D, the super-luminous stripped envelope SN~2010gx, and the SN~II-P~1999em. The inset enlarges the region around the \HeI\ $\lambda$5876 feature. Vertical solid lines mark the rest frame position of the main \HeI\ features, while the dotted, dashed, and dot-dashed lines give the rest wavelength positions of the main H and Fe\,{\sc ii} lines, along with those of the Ca\,{\sc ii} H\&K doublet. All spectra have been reddening and redshift corrected adopting the values derived in this paper and from the literature. \label{fig_spec}}
\end{figure*}

Two spectra of SN~2001gh were obtained on 2001 November 22.24 and 2001 December 20.16 using the ESO Very Large Telescope UT2, equipped with FORS1 (and the grism/filter combination 300V+GG435; Figure \ref{fig_spec}). Both observations were obtained at airmass $\sim$ 1\parcsec5 and with an average seeing around or below 1\parcsec0. The spectra were reduced following standard {\sc iraf}\footnote[8]{{\sc iraf} (Image Reduction and Analysis Facility) is distributed by the National Optical Astronomy Observatories, which are operated by the Association of Universities for Research in Astronomy, Inc. (AURA), under cooperative agreement with the National Science Foundation (NSF).} routines, including trimming, bias subtraction, and flat-field corrections. The wavelength and flux calibrations of the one-dimensional spectra were performed using arc lamp and spectrophotometric standard stars spectra, respectively, obtained at the same night and with the same instrumental configuration as the SN spectra. A check on the flux calibration was performed using the available SN photometry and, for the second spectrum of SN~2001gh, a flux correction factor was applied to match the photometric data. 

The most notable feature in the first spectrum (obtained about 10 days after discovery) is the very blue continuum, with little evidence of spectral lines. A few weak and shallow P-Cygni profiles are observed at 3889, 4471, 5015 and 5876 \AA\ (rest frame) that, in contrast with the original identification \citep{altavilla01}, we now attribute to \HeI. The lines, in particular the $\lambda$5876 feature, seems to have boxy profiles (see the insert panel in Figure \ref{fig_spec}). This is possibly an indication of the presence of an outer shell (see, for example, the boxy H$_{\alpha}$ profile in SN~1993J; e.g. \citealt{patat95}). The second spectrum (obtained approximately one month later) shows a slightly redder continuum, which is indicative of a temperature decline. Yet the \HeI\ P-Cygni lines remain the only features visible in the spectrum, which does not show clear metal features commonly seen in ordinary CC-SNe. With the revised line identifications, and the lack of spectroscopic evolution between the two epochs, we propose to reclassify this SN as a peculiar Type Ib. Considering the position of the emission peaks of the \HeI\ lines, we constrain the SN redshift to z=0.16 $\pm$ 0.01. 

In Figure \ref{fig_spec} we show a comparison of these spectra with coeval spectra of the Type Ib SN~2008D (\citealt{mazzali08,modjaz14}), the super-luminous stripped envelope SN~2010gx \citep{pasto10}, and the normal SN~II-P~1999em \citep{leonard02}. Neither the pronounced P-Cygni profile of \Ha\ characterizing spectra of SNe II (6562.8 \AA), nor the \FeII\ around 5000 \AA\ of normal SNe Ib (SN~2008D) are unequivocally detected in the spectra of SN~2001gh. The Ca\,{\sc ii} H\&K (3934, 3968 \AA) feature is instead visible in the spectra of the SN, though probably blended with HeI $\lambda$3889. In contrast with the other SNe shown in Figure \ref{fig_spec}, SN~2001gh show very little spectral evolution, and the \HeI\ lines remain the only prominent features visible in the December 20th spectrum.

The velocities, as derived from the position of the P-Cygni minimum of the most prominent \HeI\ line, the $\lambda$5876 feature\footnote[9]{The low S/N of our spectra does not allow us to properly measure the velocity of other lines.}, are $\sim$ 6600 and 6000 \kms\, respectively, for the early and late spectra of SN~2001gh. While the line velocity determined after the maximum light is comparable with that of Type Ib SNe at coeval phases, the value estimated at the earlier epoch is relatively small, being the typical line velocities of stripped-envelope SNe at this phase $\sim$ 10000 \kms; e.g. \citep{branch02}. This low velocity at peak may be an indication that SN~2001gh ejected more mass than typical SNe Ib, and/or a lower kinetic energy is released in the explosion. Additionally, we have measured a black-body temperature by fitting the continuum of our spectra. It has decreased from T$_{eff}$ $\sim$ 15000 K, as estimated in our first spectrum, to T$_{eff}$ $\sim$ 7400 K from the second one. These values are comparable with those of SN~2010gx or other super-luminous SNe at coeval epochs, as reported for example in \cite{inserra13}, but they are high in comparison with those of normal Type Ib SNe (e.g. see \citealt{stritzinger02} or \citealt{modjaz09}).

%
%
\subsection{Photometry}\label{ph}

The photometric images were reduced making use of the template subtraction technique to remove the possible contamination of the host galaxy light. Image subtraction was performed using the ISIS program \citep{alard00}. As reference images (templates) of the host galaxy, we selected those taken about two years before SN~2001gh discovery on 1999 December 12 and 1999 August 4 with the MPG/ESO 2.2~m telescope + Wide Field Imager (WFI) in $V$- and $R$-bands, respectively. After geometrical and photometric registration of the two images (target image and template), and the degradation of the image with the best seeing to match the worst one, the template was subtracted from the target image. The instrumental magnitude of the SN was measured with the point-spread function (PSF) fitting technique using the {\sc daophot} package on the subtracted image. Reference stars in the SN field were also measured using the {\sc iraf} PSF fitting routine on the original image (see Table \ref{tabla_seq}). In order to calibrate the instrumental magnitudes to a standard photometric system, we used the specific colour term equations for each instrumental configuration using average colour terms provided by the telescopes teams.

Given the high redshift estimated for this SN (see Sect. \ref{spec}), the observed $V$- and $R$-bands are respectively close to the $B$- and $V$-bands in the rest frame. The K-correction for SN~2001gh was estimated by measuring the differences between synthetic photometry in the observed and rest-frame SN spectra. Since we had only two spectra of SN~2001gh, we have assumed that the K-correction derived from the first spectrum applies to the early epochs, likewise, the K-correction derived from the second spectrum was applied for the photometry taken at the latest epoch. The K-correction for the intermediate epochs was estimated by interpolating the values derived from the two spectra. A K-correction of the order of a few tenths of a magnitude was estimated at all epochs. K-corrected magnitudes are used throughout this work (see Table \ref{tabla_ph}). Total uncertainties were computed taking into account the photometric errors and the uncertainties in the K-correction.

\begin{figure}
\centering
\includegraphics[width=\columnwidth]{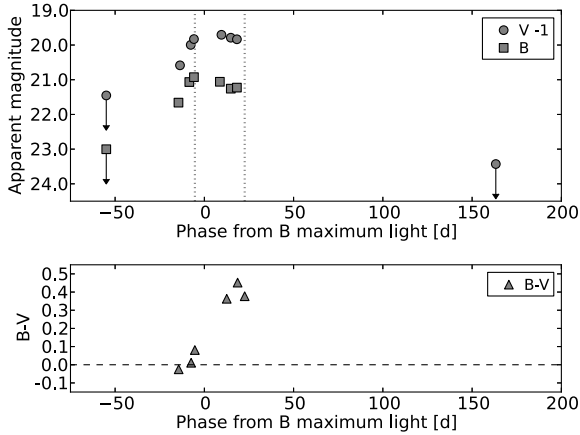} 
\caption{K-corrected $B$ and $V$ light curves of SN~2001gh (\textit{top}) and comparison of the intrinsic ($B-V$) colour
evolution of SN~2001gh and SN~2008D (\textit{bottom}). The plotted $V$-band light curve has been shifted by 1 mag. Dotted vertical lines indicate the epoch when the SN~2001gh spectra were taken. Ages are relative to $B$ maximum light in the rest frame.}
\label{fig_lightcurv}
\end{figure}

Although the photometric monitoring of SN~2001gh is rather sparse, it provides a good coverage of the SN evolution during the photospheric phase (see Figure \ref{fig_lightcurv}, top panel). Some upper limits were also measured before and after the $B$-band maximum, by introducing artificial point sources with progressively decreasing flux at the SN location. In order to estimate the epochs and magnitudes of the maximum light for the two bands, we fitted the two light curves with low-order polynomials, obtaining JD $2452242.5 \pm 1.3$ and $20.85 \pm 0.10$ mag for the $B$-band, and JD $2452247.1 \pm 1.1$ and $20.70 \pm 0.20$ mag for the $V$-band. We note that the broad light curve peaks are unusual, and not observed in normal Type Ib SNe.

In Figure \ref{fig_lightcurv} (bottom panel), the evolution of the intrinsic ($B-V$) colour of SN~2001gh in the rest frame is shown. The ($B-V$) colour of SN~2001gh  is nearly zero at early epochs, and then becomes redder until it reaches ($B-V$) $\sim$ 0.45 at 15 days after the maximum light. At this epoch, the SN seems to start a blue-ward or a flat trend. Unfortunately our data stop, preventing us to check the true behaviour of this SN. As we can also see in the figure, the overall colour evolution of SN~2001gh is different from those of normal SNe Ib such as SN~2008D \citep{modjaz09}. Additionally, the colour of SN~2001gh is bluer than SN~2008D, which is consistent with the higher temperature shown in the spectra. 

\begin{figure*}
\centering
\includegraphics[width=1.8\columnwidth]{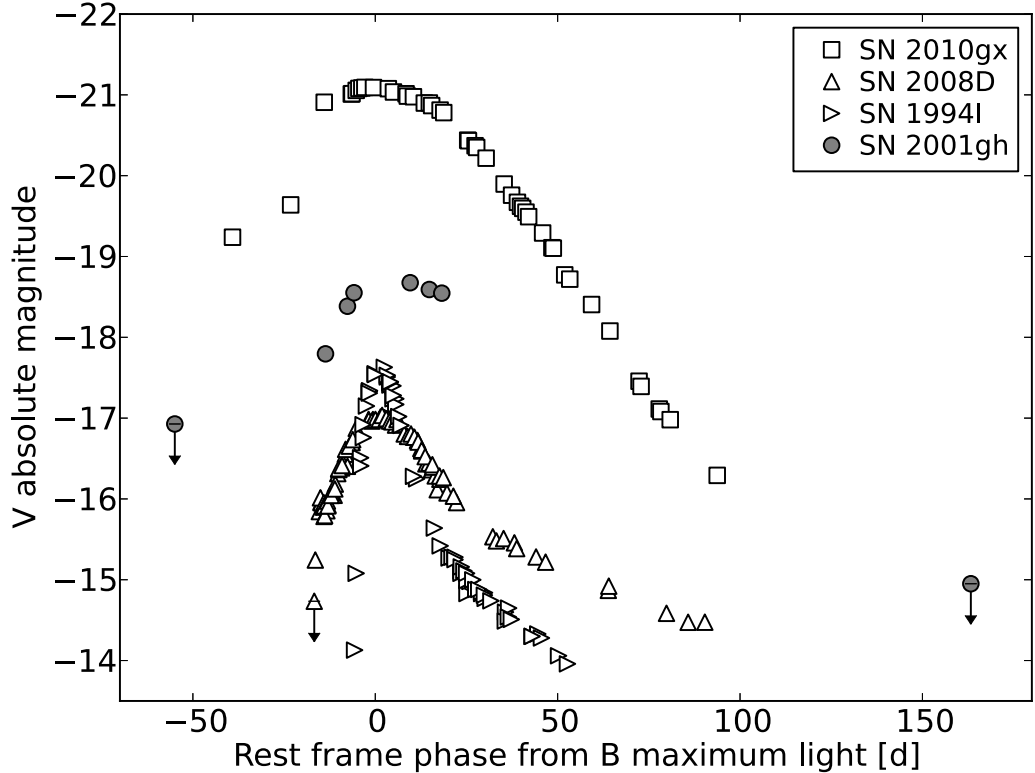} 
\caption{$V$-band absolute light curve for SN~2001gh compared with those of SNe~Ib~2008D, the super-luminous 2010gx ($B$-band absolute magnitude), and the Ic~1994I. The magnitudes of SNe~2001gh have been corrected for the assumed reddening, $E(B-V)$ of 0.016 mag, and the adopted distance modulus of 39.33 $\pm$ 0.10 mag. Distances and extinction estimates for all other SNe have been adopted from the literature. Ages are relative to $B$ maximum light in the rest frame.} \label{fig_absomag}
\end{figure*}

Figure \ref{fig_absomag} shows the rest-frame $V$-band absolute light curve of SN~2001gh, compared with those of the same stripped envelope SN sample used for the spectral comparison, viz. SNe 2008D (\citealt{mazzali08}), and 2010gx (\citealt{pasto10}), and with that of the Type Ic SN~1994I (\citealt{richmond96}). If we assume for SN~2010gx a smooth photometric evolution, its light curve peak of SN~2001gh would be very broad, with a peak luminosity that is significantly brighter than those of most CC-SNe, although much fainter than that of the super-luminous SN~2010gx.

Despite the incomplete coverage of the light curve, it seems that the SN luminosity declines quite slowly after the maximum light. This behaviour could be indicative of additional powering mechanisms, such as recombination throughout a massive He envelope or even interaction 
with circumstellar material (see discussion in Section \ref{constr}).

\begin{table}
\centering
\setlength\tabcolsep{2.5pt}
\caption{Magnitudes and associated errors of the stellar sequence used in the calibration process of SN~2001gh's photometry.}\label{tabla_seq}
\begin{tabular}{@{}lcc@{}}
\hline
Star  &  $V$  &  $R$    \\
      &  (mag) & (mag)         \\
\hline
1 & $17.11\pm0.01$ & $16.02\pm0.01$\\
2 & $15.75\pm0.01$ & $15.02\pm0.02$\\
3 & $18.73\pm0.01$ & $17.77\pm0.01$\\
4 & $22.07\pm0.01$ & $20.86\pm0.01$\\
5 & $19.81\pm0.02$ & $18.92\pm0.02$\\
6 & $20.80\pm0.10$ & $19.45\pm0.01$\\
\hline
\end{tabular}
\begin{flushleft}
Note that the photometry of SN~2001gh was calibrated in $V$- and $R$-bands before being K corrected. \\
\end{flushleft}
\end{table}
		
\begin{table}
\centering
\setlength\tabcolsep{2.5pt}
\caption{K corrected $B$- and $V$-band photometry of SN~2001gh.}\label{tabla_ph}
\begin{tabular}{@{}lcrccl@{}}
\hline
UT Date    &  JD    &Phase$^1$&  $B$  &  $V$  &  Instr.$^2$  \\
        & 2400000.0$+$ &(days)&   (mag)    &  (mag)     &          \\
\hline
04/08/99 & 51394.9 & -730.7 & - & $<23.7$ & WFI\\
03/12/99 & 51515.6 & -626.7 & $<24.5$ & - & WFI\\
26/09/01 & 52178.8 & -55.0 & $<23.0$ & $<22.5$ & WFI\\
12/11/01 & 52225.7 & -14.5 & $21.66\pm0.10$ & - & WFI\\
13/11/01 & 52226.7 & -13.6 & - & $21.58\pm0.14$ & WFI\\
19/11/01 & 52232.7 &  -8.4 & $21.07\pm0.10$ & - & WFI\\
20/11/01 & 52233.7 &  -7.6 & - & $21.00\pm0.07$ & WFI\\
22/11/01$\dagger$ & 52235.7 &  -5.8 & $20.93\pm0.07$ & $20.83\pm0.08$ & FORS1\\
09/12/01 & 52252.6 & 8.7 & $21.06\pm0.10$ & - & WFI\\
10/12/01 & 52253.6 & 9.6 & - & $20.70\pm0.31$ & WFI\\
16/12/01 & 52259.7 & 14.8 & $21.25\pm0.10$ & $20.79\pm0.07$ & EMMI\\
20/12/01$\dagger$ & 52263.7 & 18.2 & $21.23\pm0.07$ & $20.83\pm0.13$ & FORS1\\
06/06/02 & 52431.9 & 163.3 & - & $<24.4$ & FORS2\\
\hline
\end{tabular}
\begin{flushleft}
$^1$ Rest frame phases relative to the $B$ maximum (JD = $2452241.1 \pm 0.2$).\\
$^2$ WFI = ESO 2.2~m + WFI, 0.238\arcsec pix$^{-1}$; FORS1= ESO VLT UT2
+ FORS1, 0.250\arcsec pix$^{-1}$; EMMI = ESO NTT + EMMI,
0.167\arcsec pix$^{-1}$; FORS2=
ESO VLT UT1 + FORS2, 0.250\arcsec pix$^{-1}$.\\
$\dagger$ These dates are coincident with those of the spectroscopic observations.\\
\end{flushleft}
\end{table}

%
%
\section{Comparison of SN~2001gh with peculiar Type Ib SNe.}
\label{comp1bpec}

\begin{figure}
\centering
\includegraphics[width=\columnwidth]{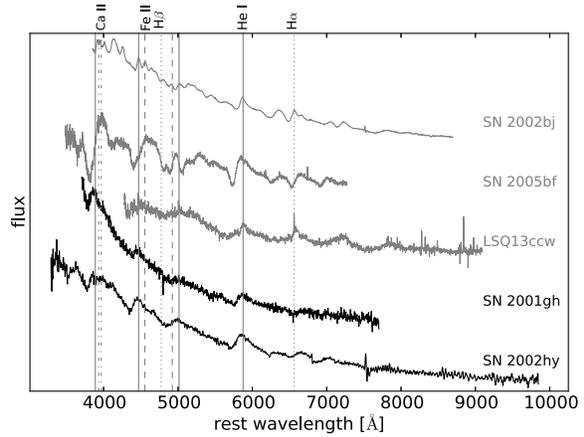} 
\caption{Comparison of the early optical spectrum of SN~2001gh with those of the peculiar Type Ib SNe~2002bj, 2002hy, 2005bf, and the Ibn LSQ13ccw at epochs between -5 and 11 days from the maximum light. Vertical solid lines mark the rest frame position of the main \HeI\ features, while the dotted, dashed, and dot-dashed lines give the rest wavelength positions of the main H and Fe\,{\sc ii} features, along with those of the Ca\,{\sc ii}~H\&K doublet. All spectra have been reddening and redshift corrected assuming the values adopted in this work and from the literature. 
\label{fig_spec02hy}}
\end{figure}

To place SN 2001gh in the context of the Type Ib SN zoo, we compare in Figure \ref{fig_spec02hy} our earliest spectrum of SN~2001gh with spectra of a sample of peculiar Ib SNe at similar epochs, viz. SNe~2002bj (a peculiar, fast-evolving SN Ib; \citealt{poznanski10}), 2002hy (a relatively narrow-lined SN Ib from the Padova-Asiago SN archive, cfr. appendix \ref{sn02hy}), 2005bf (a Type Ib event with a double-peaked light curve; \citealt{folatelli06,modjaz14}), and the unusual Type Ibn LSQ13ccw \citep{pastorello15}. Among them, an object that shares some spectroscopic similarity with SN 2001gh is SN~2002hy. The spectra of these two SNe are blue, with \HeI\ lines showing evident P-Cygni profiles, such as $\lambda$3889, $\lambda$4471, $\lambda$5015 and $\lambda$5876. Weak Ca\,{\sc ii} H\&K ($\lambda\lambda$3934,3968) features are also detected in the spectra of the two objects. 

The most remarkable difference is that the P-Cygni absorption at around 5870 \AA\ is more blue-shifted in the spectrum of SN~2002hy than in SN~2001gh. In fact, the velocity of the ejecta of SN~2002hy, as derived from the position of the P-Cygni minimum of the prominent \HeI\ $\lambda$5876 line, is $\sim$ 8000 \kms, higher than that estimated for SN~2001gh in the early epoch spectrum. In addition, SN~2002hy shows a cooler spectrum with a measured black-body temperature of $\sim$10000 K, versus $\sim$15000 K estimated from the SN~2001gh one. On the other hand, the properties of the spectra of SNe~2001gh and 2002hy are significantly different from those of any other SN considered in Figure \ref{fig_spec02hy}.

\begin{figure}
\centering
\includegraphics[width=\columnwidth]{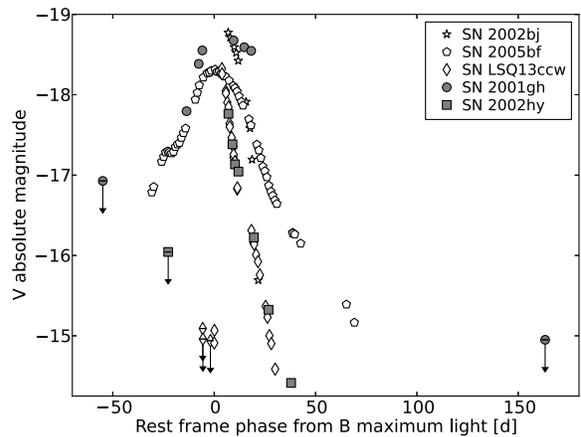} 
\caption{$V$-band absolute light curve for SN~2001gh compared with those of the peculiar SNe~Ib~2002bj, 2002hy ($R$-band absolute magnitudes), 2005bf, and the Type Ibn~LSQ13ccw (sloan $r$-band converted into Johnson-Cousins $R$-band). The magnitudes of SNe~2001gh and 2002hy have been corrected for the reddening values adopted in this paper, $E(B-V)$ of 0.016 and 0.141 mag, respectively, and accounting for their adopted distance moduli of $39.33 \pm 0.10$ and $33.61 \pm 0.20$ mag. Distances and extinction estimates for all other SNe have been taken from the literature. Ages are relative to the rest-frame $B$ maximum light.  
} \label{fig_absomag02hy}
\end{figure}

In Figure \ref{fig_absomag02hy} we compare the $V$-band absolute light curve of the same sample of objects shown in Figure \ref{fig_spec02hy}. We may immediately note the large heterogeneity in these light curves. In particular, SN~2001gh is brighter than any other but SN~2002bj, and shows the slowest post peak decline. \\

Based on the comparison described above, we consider SN~2001gh an unprecedented event. The object shares spectroscopic similarity with SN~2002hy but their light curves are very different. If the spectra of the two SNe may suggest similar gas properties at the photosphere, their photometric evolutions are also indicative that the progenitor parameters and/or the explosion energy may have been significantly different. In this context, we note the remarkable photometric similarity of SN~2002hy with the light curve of the Type Ibn SN LSQ13ccw (Figure \ref{fig_absomag02hy}), whose luminous peak and fast photometric evolution were explained in terms of interaction between the SN ejecta and He-rich CSM \citep{pastorello15}. Although SN~2002hy does not show the narrow He I emission from the CSM lines observed in Type Ibn SNe such as LSQ13ccw, we cannot rule out that a similar mechanism may also power the light curve of SN~2002hy. In this case, the lack of narrow lines from the photoionised gas may be explained with a very high-density CSM (e.g. see \citealt{blinnikov10,chevalier11,moriya13}).

Is this scenario suitable also for SN~2001gh? SN~2001gh has a luminous and broad light curve peak which could indicate a moderate mass of $^{56}$Ni, but the non-detection at $\sim$165 days after maximum light suggests that the ejected $^{56}$Ni mass cannot be the primary source powering its light curve.
In fact, we will see in section \ref{constr} that the ejecta-CSM scenario suitable for Type Ibn SNe can provide a promising explanation also for SN~2001gh-like events, although alternative mechanisms will be discussed.

%
%
\section{Constraining the powering mechanism of SN~2001gh}
\label{constr}

It is clear from the data  illustrated in the previous sections that SN~2001gh is a peculiar SN Ib with unprecedented observational properties, showing some spectroscopic similarity with the usual SN Ib 2002hy. However, the latter has a much faster evolving light curve, similar to those of some Type Ibn SNe, such as SN 1999cq, 2000er and LSQ13ccw (\citealt{matheson00,pastorello08,pastorello15}), whilst SN~2001gh has a broad light curve with a slow post-peak decline, reminiscent of some SLSNe (e.g. \citealt{quimby11}). From the discussion in section \ref{comp1bpec}, we suggested that the observed properties of SNe~2001gh and 2002hy are not primarily driven by the radioactive decays. A way to verify this claim is through data modelling.

\begin{figure*}
\begin{center}
\centering
\includegraphics[width=2.2\columnwidth]{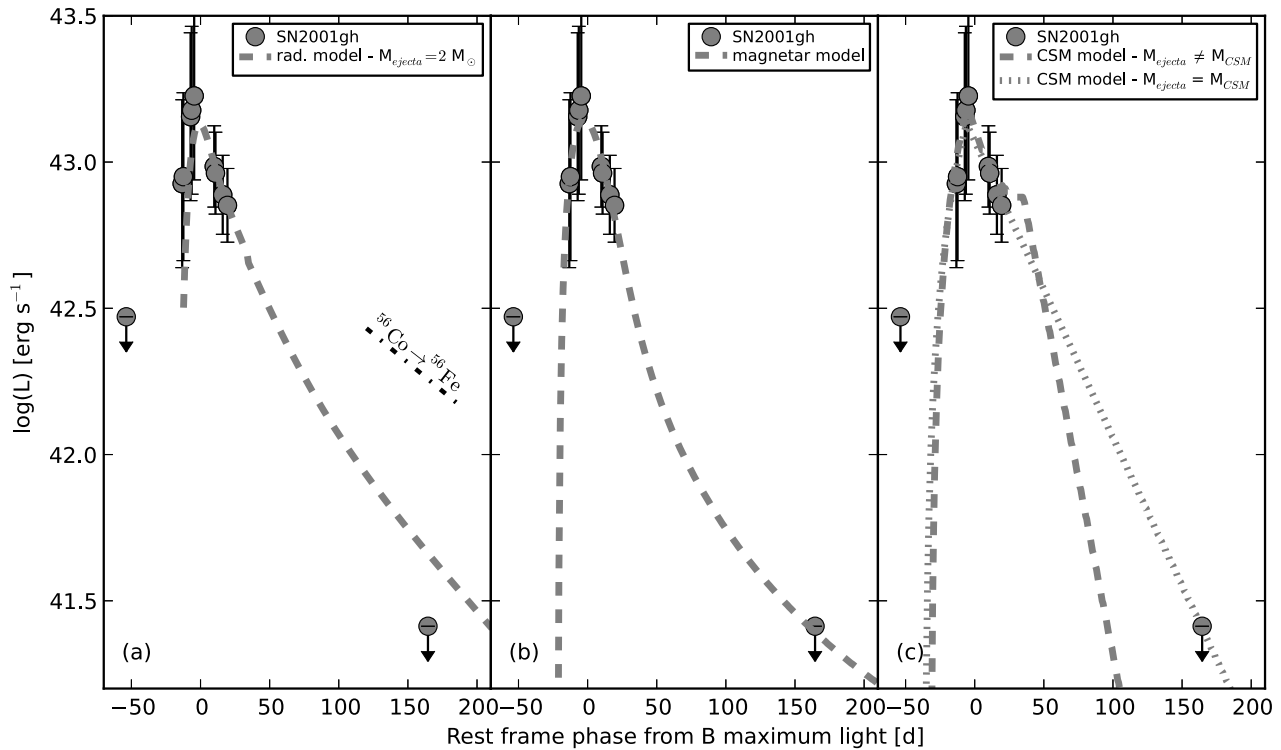}
\caption{Comparison between the bolometric light curve of SN~2001gh (see the text for more details) and the resulting best fits of the model of {\it (a panel)} radioactive decay for M$_{ejecta} = 2$ M$_{\odot}$ (dashed line),  {\it (b panel)} magnetar (dashed line), and  {\it (c panel)} CSM interaction for M$_{ejecta} \neq$ M$_{CSM}$ (dashed line) and M$_{ejecta} =$ M$_{CSM}$ (dotted line) explosion scenarios. }
\label{fig_modelling}
\end{center}
\end{figure*}

In order to constrain the parameters of the explosion of SN~2001gh and, hence, the progenitor's scenario, we have estimated an approximated bolometric light curve for SN~2001gh. Since only two photometric bands were available for the light curve of SN~2001gh, the estimate of the bolometric luminosity required some assumptions and a few operational steps. Firstly, we derived the BV light curve by calculating the luminosity from the extinction-corrected apparent magnitudes in these bands and for our adopted distance. Then, since the SN~2001gh spectra are practically featureless, we computed the total luminosity from 900 to 26000 \AA\ of two black bodies with the temperatures derived for each of the two spectra of SN~2001gh (see section \ref{spec}). We estimated the bolometric corrections at these two epochs, and applied them to the BV {\it pseudo}-bolometric luminosities following a similar criteria than the one applied for the K-correction, i.e., we assumed the same correction derived from the first spectrum for the early epochs, the correction derived from the second spectrum for the late epochs, and for the intermediate epochs, we integrated the total flux of black bodies with intermediate temperatures assuming a smooth evolution of the continuum temperature. The errors in the bolometric luminosity account for the uncertainties in the estimates of the distance, extinction, apparent magnitudes and black body temperature. The inferred bolometric light curve is shown in Figure \ref{fig_modelling}. The luminosity at maximum light was estimated by fitting the light curve with low order polynomials, resulting in $L_{max} \approx (1.7\pm0.9)$x$10^{43}$ erg s$^{-1}$.\\

Since the limited data set available does not allow us to robustly constrain the explosion scenario, we attempt to compute the explosion parameters and the nature of the progenitor star by considering three alternative scenarios for explaining the observables of SN~2001gh, in particular its bolometric light curve: 

\begin{itemize}

\item
{\bf Radioactively-powered SN scenario --} In our first model, we assume that the luminosity of SN~2001gh is mostly  powered by the energy released in the radioactive decay chain $^{56}$Ni - $^{56}$Co - $^{56}$Fe. In this case, we compare the light curve of SN~2001gh with the output light curves obtained with a toy model based on a simplified treatment of \cite{arnett82} and described in \cite{valenti08}\footnote[10]{This model uses the analytic approximations of \cite{arnett82}, accounting for the typo correction reported in \cite{arnett96} (see also \citealt{wheeler15})}. This model, which has been successfully used for modelling a number of stripped envelope SNe (e.g. \citealt{moralesgaroffolo14}), divides the SN evolution into a photospheric and nebular phase, assuming expanding ejecta with spherical symmetry. During the photospheric phase, the model includes the energy produced by the radioactive decays. During the nebular phase, instead it assumes that the SN luminosity is powered only by the energy deposition of $\gamma$-rays from the \Cofs\ decay. Accounting the photospheric velocities derived before (see section \ref{spec}), and the host galaxy distance and extinction values as discussed in section \ref{ph}, we obtain a fairly good match with the bolometric light curve\footnote[11]{In order to fit the late phases of the SN, we considered in all cases the last upper detection limit as a genuine detection. This is intended to test the most conservative scenario.} adopting a kinetic energy between $0.6$x$10^{51}$ and $3.5$x$10^{51}$ erg, ejecta mass of about $2$ M$_{\odot}$, and \Nifs\ mass $\sim 0.6$ M$_{\odot}$ (see panel {\it a} of Figure \ref{fig_modelling}). Although this model reproduces fairly well the obtained bolometric light curve, the small ejecta mass, composed by about 30\% in mass of nickel, does not easily accommodate in a typical CC-SN explosion scenario. Similar M(\Nifs)/M$_{ejecta}$ ratio can be found, in fact, in some rare luminous Type Ia SNe (e.g. \citealt{taubenberger13}). However, in contrast with what expected in a thermonuclear SN explosion scenario, there is no evidence of strong features of iron-group elements in the spectra of SN~2001gh. 
In addition, the total ejected mass here derived is comparable with the average value for a normal Type Ib SN (see e.g. \citealt{drout11}). However,  to reproduce the broad luminosity peak and the slow spectral evolution of SN~2001gh, one would have expected a larger ejected mass.
In summary, for different reasons, the model illustrated above does not support a scenario where the SN~2001gh luminosity is primarily powered by radioactive decays.

\item 
{\bf Magnetar spin-down scenario -- } An alternative possibility to explain the properties of SN~2001gh is that the energy powering its luminosity comes from the spin-down of a newly formed magnetar, instead of the canonical radioactive decays (see e.g. \citealt{kasen10}). We explore this possibility using a magnetar model as presented and discussed by \cite{inserra13} and \cite{nicholl14}. The observed light curve of SN~2001gh is compared to a grid of magnetar-powered synthetic SN light curves, through $\chi^2$ minimization. The observed light curve satisfactorily fits a model obtained with ejected mass of $\sim$ 3.4 M$_{\odot}$, a magnetic field  B $\approx$ $11$x$10^{14}$ G, and a spinning period of $P_i \approx$ 12 ms (see panel {\it b} of Figure \ref{fig_modelling}). This B field is high but in combination with the $P_i$ estimated, this magnetar could produce a luminosity peak of $10^{43}$ erg s$^{-1}$ \citep{kasen10}, which is in agreement with the maximum luminosity derived with our bolometric light curve. However, this scenario alone does not clearly explain the slow spectroscopic evolution of SN~2001gh or the boxy profiles of the He I lines.

\item
{\bf Ejecta-CSM interaction scenario --}
The radioactively-powered models predict a faster evolution of $v_{ph}$, which is not observed in the available spectra of SN~2001gh. If we consider another observational constraint from the spectroscopy of SN~2001gh, i.e. some evidence of boxy profile in the \HeI\ lines, an alternative interpretation could be offered, according to which SN~2001gh is a more canonical stripped envelope SN exploded in a dense, He-rich circumstellar shell. This CSM was produced through mass-loss episodes occurred prior to the SN explosion. To support this suggestion, we modelled the bolometric light curve of SN~2001gh using a semi-analytic code which accounts for ejecta-CSM collision as the main powering source (panel {\it c} of Figure \ref{fig_modelling}; see \citealt{chatzopoulos12} or \citealt{nicholl14}, and references therein). To obtain a good match between the light curve output of this interaction model and the observed peak of luminosity of SN~2001gh would require CSM/ejecta mass ratios in the range $0.1$-$0.2$ to $1$, assuming always a modest kinetic energy and an opacity of 0.2 cm$^2$ g$^{-1}$. For example, in Figure \ref{fig_modelling} we plotted two of the best interaction-powered models with $\sim$ 7 M$_{\odot}$ of ejecta and $\sim$1 M$_{\odot}$ of CSM (kinetic energy $\sim 0.25$x$10^{51}$ erg), and with M$_{ejecta} =$ M$_{CSM} \approx 2.4$ M$_{\odot}$ (kinetic energy $\sim 0.1$x$10^{51}$ erg). 

\end{itemize}

Therefore, accounting for  the comparison between the observables of SN~2001gh (in particular its bolometric light curve), and the expectations from the models, we consider more plausible that the luminosity of this SN has been mostly powered by the interaction of the ejecta with an opaque CSM, probably ejected by the progenitor star years prior to the SN explosion. Accordingly, this scenario is supported by the \HeI\ line profiles observed in the SN~2001gh spectra. The low P-Cygni velocities and the modest velocity gradient estimated from the spectra would indicate that the lines formed in shocked CSM rather than in the SN ejecta.
In summary, we cannot provide conclusive arguments to rule out some the alternative scenarios, in particular the magnetar spin-down powering mechanism. On the other hand, the radioactive decays alone cannot provide a reasonable explanation for the properties of this peculiar Ib SNe.

%
%
\section{Summary}
\label{sum}

In this paper we have presented spectroscopic and photometric observations of the peculiar SN~Ib~2001gh, discovered at a redshift z=0.16 by the STRESS survey. Assuming a smooth evolution, SN~2001gh shows an unusual broad and delayed light curve peak, reaching $-18.55$ mag at $B$ maximum light. On the other hand, the two spectra available for SN~2001gh, taken at around 30 days from each other, show a quasi featureless blue continuum, with a few weak P-Cygni lines that we attribute to \HeI\ at a velocity of 6000-7000 \kms. 

Three possible explosion scenarios have been investigated to explain the observables of SN~2001gh: according to the first one, the SN luminosity may arise from radiated energy from the $^{56}$Ni to $^{56}$Co to $^{56}$Fe decay chain, where the explosion ejected a total mass of $\sim 2$ M$_{\odot}$ (about $0.6$ M$_{\odot}$ of which being \Nifs); in the second scenario, the SN luminosity is powered by a magnetar with a magnetic field B $\approx 11$x$10^{14}$ G and a spinning period of about 12 ms; the third scenario suggests a more canonical SN Ib embedded in a dense, circumstellar shell. Weighting up all the observables and the resulting models of SN~2001gh, and accounting for the spectroscopic similarity with the peculiar SN~2002hy (and more marginally with Type Ibn SNe), we believe that the scenario with the peak SN luminosity being powered mostly by the interaction of the ejecta with an optically thick shell is the most reliable. Unfortunately, this dense CSM does not allow us to identify the true chemical composition of SN~2001gh, which is key information to unveil the nature of the progenitor star.

SN~2001gh is an unprecedented object. Assuming that this event is representative of a distinct class, we can derive an indicative estimate the rate of occurrence by using the recipes described in \cite{botticella08}. The main uncertainty in the calculation is the incomplete coverage of the light curve that we address by exploring two extreme cases for the evolution after maximum, either a decline similar to SNe II-Plateau such as SN~2004et \citep{maguire10}, or to SNe II-Linear like SN~1979C \citep{barbon82}. In both cases, the light curve is scaled to the observed magnitude at maximum of SN~2001gh. With this prescription, and considering the observing log and detection efficiency of the STRESS survey, we derive a rate for SN~2001gh-like events to be in the range $1$-$4$x$10^{-6}$ yr$^{-1}$Mpc$^{-3}$, which is 30 to 100 times smaller that the overall rate of core collapse SNe at reshift $z \sim 0.2$.

The outstanding properties of SN~2001gh are an indication that our current knowledge on the mechanisms producing the variety of observed parameters of CC SNe is still incomplete. Object like SN~2001gh are intrinsically rare, and panoramic surveys and spectroscopic classification programs should help to discover further members of this family.

%
%

\bigskip

\noindent {\bf ACKNOWLEDGMENTS}

We are grateful to M. Hamuy, G. Pignata and J.~B.~Vanssay for the SN~2002hy data. 
NER acknowledges the support from the European Union Seventh Framework Programme (FP7/2007-2013) under grant agreement n. 267251 ``€œAstronomy Fellowships in Italy" (AstroFIt). NER, AP, SB and MT  are partially supported by the PRIN-INAF 2014 with the project ``Transient Universe: unveiling new types of stellar explosions with PESSTO".
This work is based on observations made with ESO Telescopes at the La Silla and Paranal Observatories under programme ID 064.H-0390, 067.A-0497, 068.A-0443, 068.D-0273, 069.A-0312, 070.D-0721. This work has made use of the NASA/IPAC Extragalactic Database (NED), which is operated by the Jet Propulsion Laboratory, California Institute of Technology, under contract with the National Aeronautics and Space Administration.
\noindent

\bsp

\bibliographystyle{mn2e}
\bibliography{bibl}
\label{lastpage} 

\appendix

\section{SN~2002hy}
\label{sn02hy}

\begin{figure}
\centering
\includegraphics[width=0.9\columnwidth]{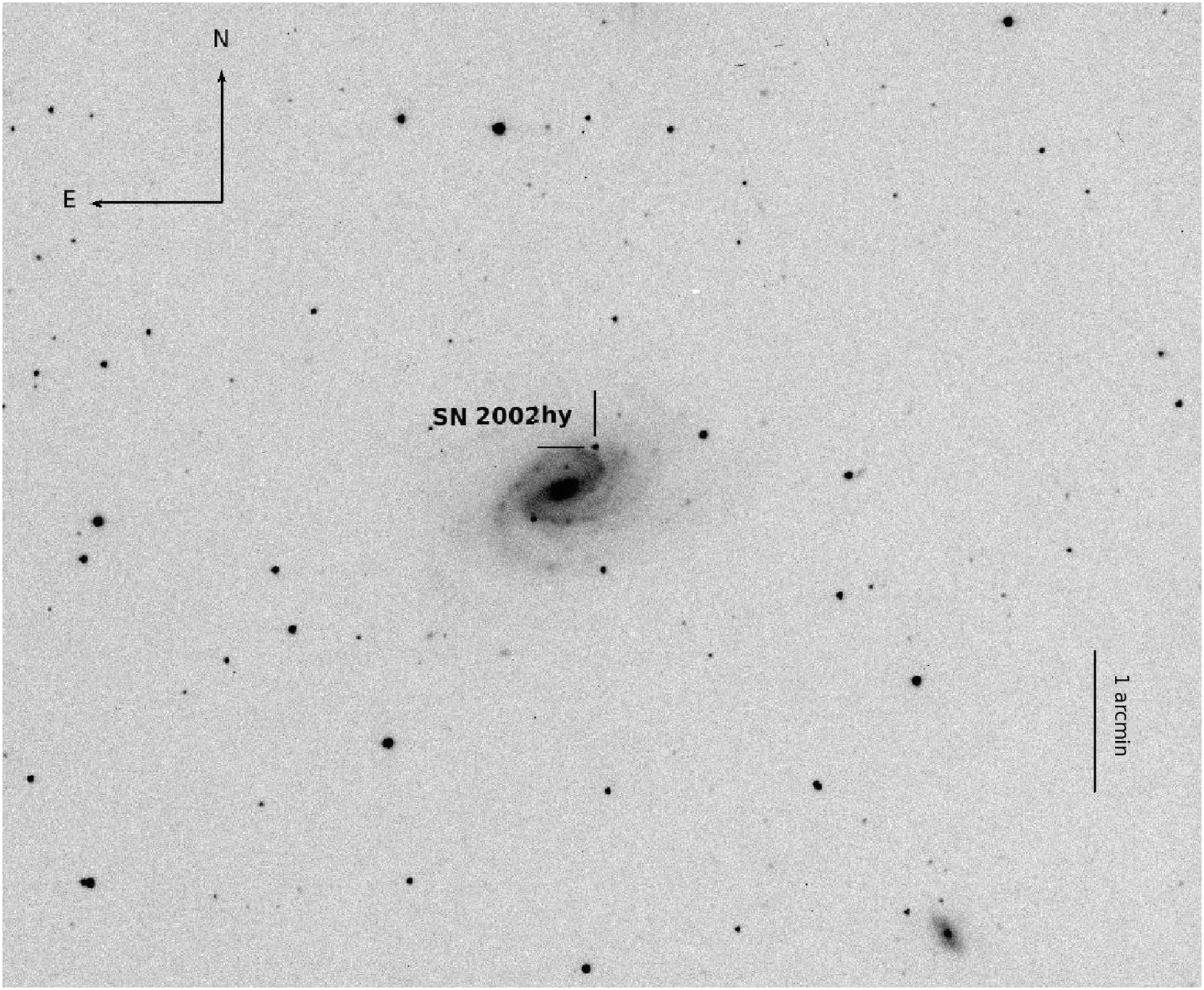} 
\caption{$I$-band image of SN~2002hy field taken with LCO 40 inch + CCD on 2002 November 18. The SN position is marked. }
\label{fig_FC02hy}
\end{figure}

SN~2002hy was discovered by \cite{monard02} on 2002 November 12.10 UT in NGC~3464. Its coordinates are: $\alpha = 10^{\rm h} 54^{\rm m} 39{\fs}18, \delta = -21\degr 03\arcmin 41{\farcs}2$ (J2000.0; Figure \ref{fig_FC02hy}). The object was initially spectroscopically classified as a peculiar Type Ib a few days after the explosion. The spectrum, in fact, was dominated by a  blue continuum with super-imposed  strong \HeI\ lines \citep{benetti02,harutyunyan08}. We note that a few months prior to the discovery of SN~2002hy, on 2002 January 21.4 UT, another object \citep[SN~2002J ,][]{weisz02} was observed in the same host galaxy, and classified as a Type Ic near maximum light \citep{cappellaro02}. 

NGC~3464 is a rather luminous (M$_B \sim -21.4$ mag) barred spiral galaxy, with a distance modulus of $\mu = 33.61 \pm 0.20$ mag\footnote[12]{NED, NASA/IPAC Extragalactic Database; http://nedwww.ipac.caltech.edu/.}. We also assume a total foreground extinction $E(B-V) = 0.141 \pm 0.030$ mag, including a contribution form the Galactic reddening of $E(B-V)_G = 0.048$ mag \citep{schlafly11}, and an additional contribution of $E(B-V)_{NGC~3464} = 0.093$ inside the host galaxy. The latter was estimated from the equivalent width of the narrow Na\,{\sc id} interstellar feature and the relation given by \cite{poznanski12}.

To estimate the explosion date of SN~2002hy, we rely on the similarity of its $R$-band light curve with that of LSQ13ccw (Figure \ref{fig_absomag02hy}; \citealt{pastorello15}). Being the explosion date of LSQ13ccw well constrained, and assuming a similar rising time for the two SNe, we estimate the explosion of SN~2002hy to occur on JD = 2452582.7 $^{+1}_{-5}$ (i.e., 2002 November 04.20 UT). This date is consistent with the discovery date on 2002 November 12 (JD = 2452590.6), and the last non-detection of the SN on 2002 October 13 (JD = 2452560.6). Deep images of SN~2002J taken on 2002 January 25 with the ESO New Technology Telescope (NTT) and EMMI (ESO Multi-Mode Instrument) were also checked, showing no source detected at the SN position.

One spectrum of SN~2002hy was taken at the ESO Very Large Telescope 3.6-m telescope with EFOSC2 (Faint Object Spectrograph and Camera - v.2; using grism 11 and 12) on 2002 November 15.33 (JD = 2452593.83; Figure \ref{fig_spec02hy}), at phase  $\sim$ 11 days after the presumed explosion epoch. The spectrum was reduced following the prescriptions illustrated in section \ref{spec}. 

The photometric data of SN~2002hy were reduced as detailed in section \ref{ph}. For the amateur data, an image of NGC~3464 taken on 2004 May 10 with a 0.3m telescope (+ SBIG ST-7 CCD) was used as a template to subtract the host galaxy contamination. The apparent magnitudes of the SN in the $R$-band and the associated errors are reported in Table \ref{tabla_ph02hy}. According to the comparison in Figure \ref{fig_absomag02hy}, SN~2002hy seems a photometric twin SN of the Type Ibn LSQ13ccw, having a fast average decline after maximum of $\gamma_R \sim$ 10.6 $\pm$ 0.5 mag/100d, comparable with those of the Type Ibn (between 8 and 13 mag/100d; e.g. see \citealt{pastorello15}). Therefore, considering also the information on the photometric evolution, we propose to re-classify SN~2002hy as a transitional Type Ibn/Ib SN, although  with  He~I spectral lines which are broader than observed in other Type Ibn SNe (see e.g. Figure \ref{fig_spec02hy}). This sub-class of transitional events will be discussed in a forthcoming paper (Pastorello et al. in preparation).\\

\begin{table}
 \centering
  \setlength\tabcolsep{2.5pt}
   \caption{$R$-band photometry of SN~2002hy.}\label{tabla_ph02hy}
  \begin{tabular}{@{}lcccl@{}}
  \hline
 UT Date    &  JD    & Phase$^1$ & $R$  &  Instr.  \\
        & 2400000.0$+$    &  (days) &   &          \\
 \hline
13/10/02 & 52560.6 & -22.1 & $<18.0$         &    L.A.G.Monard\\
12/11/02 & 52590.6 & 7.9 & $16.28\pm0.18$ &    L.A.G.Monard\\
14/11/02 & 52592.6 & 9.8 & $16.66\pm0.17$ &    L.A.G.Monard\\
15/11/02$\dagger$ & 52593.9 & 11.1 & $16.91\pm0.13$ &    ESO3.6m\\
17/11/02 & 52595.6 & 12.8 & $17.00\pm0.14$ &    L.A.G.Monard\\
24/11/02 & 52603.3 & 20.5 & $17.82\pm0.23$ &    AAT\\
02/12/02 & 52610.6 & 27.8 & $18.72\pm0.34$ &    L.A.G.Monard\\
13/12/02 & 52621.8 & 39.0 & $19.63\pm0.29$ &    J.B.Vanssay\\
\hline
\end{tabular}
\begin{flushleft}
$^1$ Relative to the explosion date (JD = 2452582.7).\\
L.A.G.Monard = 0.3m Telescope + SBIG ST-7, 0.91\arcsec pix$^{-1}$; ESO3.6m = ESO 3.6m Telescope + EFOSC2, 0.33\arcsec pix$^{-1}$.; AAT = Anglo-Australian Telescope + WFI, 0.25\arcsec pix$^{-1}$; J.B.Vanssay =  0.28m Telescope + KAF-400, 0.85\arcsec pix$^{-1}$\\
$\dagger$ Date in which the spectrum was taken.\\
\end{flushleft}
\end{table}

\end{document}